\documentclass[a4paper,11pt]{article}
\usepackage{pos}
\newcommand{\KK}{KK}

\title{KKMC-hh for Precision Electroweak Phenomenology at the LHC}
\ShortTitle{KKMC-hh for Precision EW Phenomenology at the LHC}

\author*[a]{Scott A. Yost}
\author[a]{M. Dittrich}
\author[b]{S. Jadach}
\author[c]{B.F.L. Ward}
\author[b]{Z. W{\c a}s}

\affiliation[a]{The Citadel,\\
  171 Moultrie St., 
  Charleston, SC 29409, USA}

\affiliation[b]{Institute of Nuclear Physics, Polish Academy of Sciences,\\
  ul.\ Radzikowskiego 152, 31-342 Krak\'ow, Poland}

\affiliation[c]{Baylor University,\\
  Waco, TX 75798, USA}

\emailAdd{scott.yost@citadel.edu}
\emailAdd{mdittric@citadel.edu}
\emailAdd{stanislaw.jadach@cern.ch}
\emailAdd{bfl\_ward@baylor.edu}
\emailAdd{z.was@cern.ch}

\abstract{
We describe the program {\KK}MC-hh, which calculates Z boson processes in hadronic
collisions using coherent exclusive exponentiation (CEEX) with exact 
second-order photonic corrections at next-to-leading log and first-order weak
vertex corrections, including initial and final state photonic radiation and 
initial-final interference.  We describe current applications to precision 
forward-backward asymmetry calculations for the measurement of the electroweak
mixing angle at the LHC. 
\\[1cm]
\begin{center}
BU-HEPP-20-08
\end{center}
}

\FullConference{%
  40th International Conference on High Energy physics - ICHEP2020\\
  July 28 - August 6, 2020\\
  Prague, Czech Republic (virtual meeting)
}

\begin{document}
\maketitle

\section{Introduction and Program Structure}
{\KK}MC-hh\cite{kkmchh} is a precision MC generator for Z production and decay in high-energy
proton collisions. It is based on {\KK}MC,\cite{Jadach:1999vf} which was originally developed for 
precision Z boson phenomenology in $e^+e^-$ collisions, including exponentiated
multiple photon effects: \hbox{$e^+e^-\rightarrow Z\rightarrow f{\overline f} + n\gamma$}
including exact ${\cal O}(\alpha)$ and ${\cal O}(\alpha^2L)$ initial-state 
radiation (ISR), final-state radiation (FSR), and initial-final interference
(IFI). ($L$ is an appropriate ``big logarithm'' for the process, $\ln(p^2/m^2)$
for a relevant mass.) 
Order $\alpha$ electro-weak matrix element corrections are included via 
an independent DIZET module, originally version 6.21\cite{zfitter6:1999}, 
but recently upgraded to
version 6.45\cite{Arbuzov:2005ma,Arbuzov:2020}. Collision energies up to 1 TeV are supported. 
The LEP2 precision
tag was $0.2\%$. Beginning with version 4.22, {\KK}MC also includes support for
parton-level collisions of quarks.\cite{Jadach:2013aha}

An adaptive MC, FOAM,\cite{FOAM} underlies the low-level event generation. 
The FOAM grid is set up during an initial exploratory
phase, creating a crude MC distribution that includes the PDF factors and a 
crude YFS form-factor for the ISR photon radiation. 

{\KK}MC generates multiple-photon radiation using one of two modes of resummation.
EEX mode (exclusive exponentiation) is based on YFS soft photon resummation,\cite{yfs}
implemented at the cross-section level, while CEEX mode (coherent exclusive 
exponentiation)\cite{Jadach:2000ir} is an amplitude-level adaptation of YFS
resummation. IFI enters
naturally when an amplitude including exponentiated ISR and FSR factors is 
squared.  See Ref. \cite{JadachYost2019} for a recent study of IFI in the CEEX framework.

The events generated by {\KK}MC-hh may be showered externally by exporting them in an
LHE-format event file,\cite{lhe-format} or by running a built-in HERWIG6.5\cite{HERWIG} shower.  

\section{Photonic Radiative Corrections}

{\KK}MC-hh takes 
an ab-initio calculation of photon ISR at the Feynman diagram level, with 
exponentiation. This is in contrast to a more
traditional approach of factorizing the collinear ISR into the parton 
distribution functions.  

A QED-corrected PDF can account for the shift in quark moment due to photonic
ISR to the extent that the observable is sufficiently inclusive to average 
over any transverse momentum. In particular, it is reasonable to expect that
{\KK}MC-hh should show good agreement with a QED-corrected PDF for distributions
such as the invariant-mass distribution of the final leptons, in the absence
of individual lepton cuts. However, {\KK}MC-hh can go beyond the PDF approximation and account
for the transverse momentum effects from the ISR that would be missed in a strictly 
collinear representation that effectively confines the ISR to the protons.

Fig. 1(a) shows the ratio of the invariant mass ($M_{ll}$) distribution
with ISR turned on in {\KK}MC-hh to the distribution with ISR off (red), compared
the same ratio calculated with ISR off in {\KK}MC-hh, but using the LuxQED version of NNPDF3.1
instead of the standard version.  The results are from a $10^9$ event muon sample 
generated without a QCD shower or FSR and with no cuts on the muon momenta, 
using current quark masses from the PDG,\cite{PDG2018} for 8 TeV proton CM energy. 

Both methods of accounting for ISR lead to
a shift of about $-0.5\%$, and the distributions largely overlap to the 
statistical limits of the sample, suggesting a high degree of consistency
between the two approaches in a case where agreement is expected. The shift in the total 
cross section for $60\ {\rm GeV} < M_{ll} < 120\ {\rm GeV}$ was $-0.524\pm 0.004\%$ when turning
on ISR in {\KK}MC-hh, and $-0.624\pm 0.002\%$ when switching to the NNPDF3.1-LuxQED, so the
two ways of accounting for ISR give cross sections in agreement to $0.1\%$. The shift in the 
average $M_{ll}$ was $-0.10\pm 0.27\ {\rm MeV}$ when turning on ISR in {\KK}MC-hh, and $-0.40
\pm 0.27\ {\rm MeV}$ when switching to a LuxQED PDF instead.

Fig. 1(b) shows a similar comparison for the rapidity distribution $Y_{ll}$
for the same events. Both {\KK}MC and the LuxQED version of NNPDF3.1 lead to 
a shift in the rapidity distribution on the order of $-0.5\%$, but the 
shape is somewhat different, crossing at $Y=2.5$.

\begin{figure}[ht]
\centering
\includegraphics[width=0.49\textwidth]{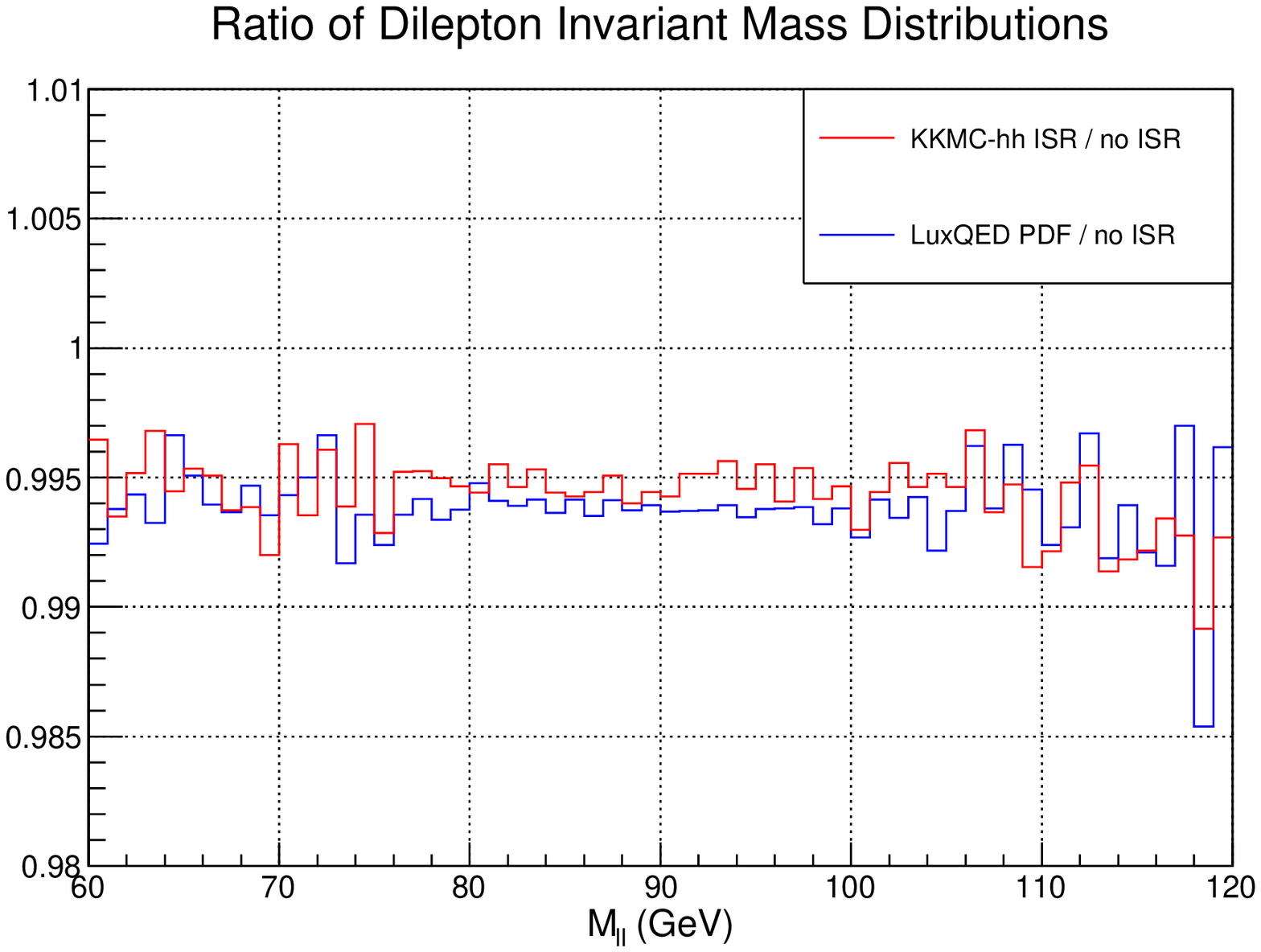}
\includegraphics[width=0.49\textwidth]{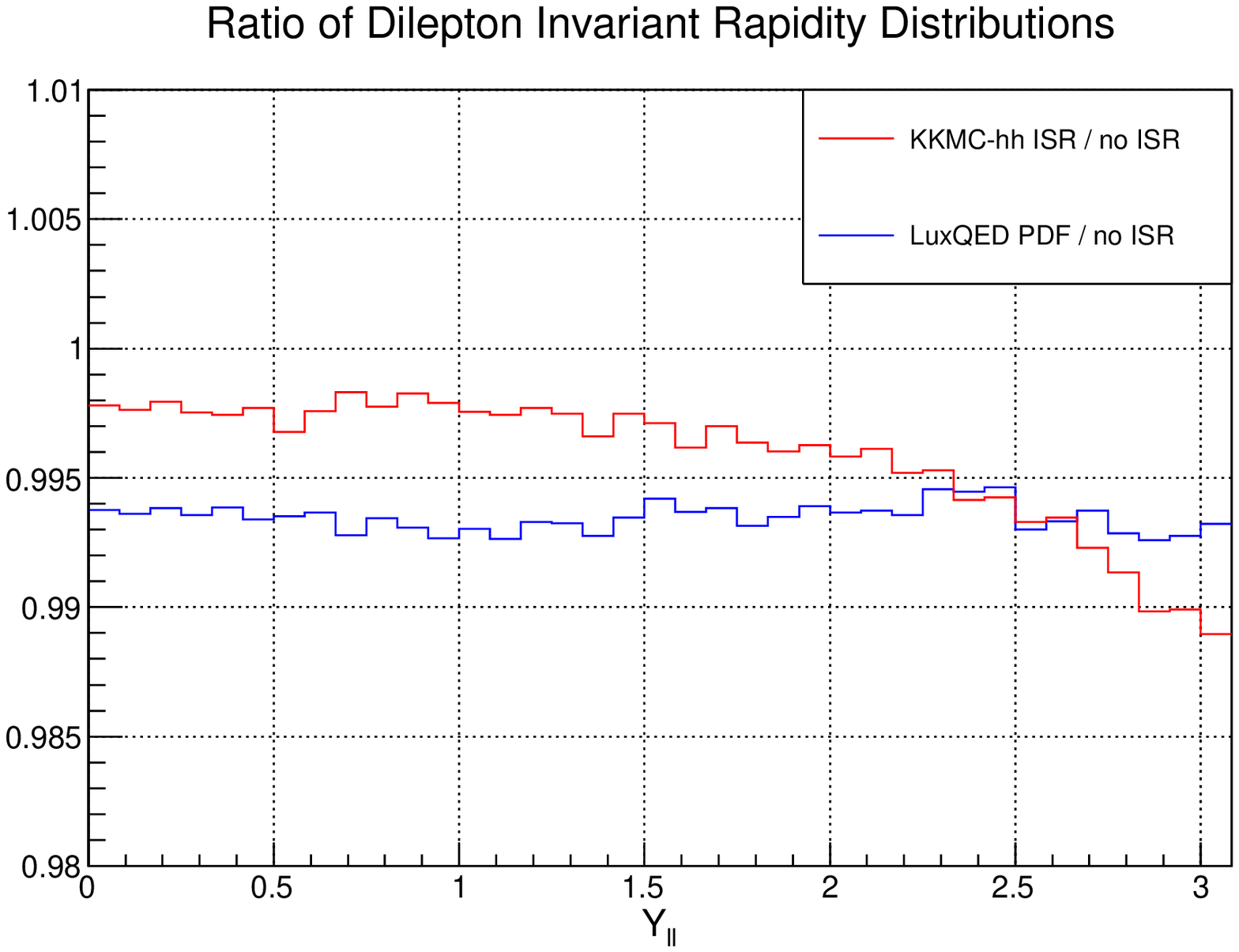}
\caption{Comparison of the ratios of the invariant mass ($M_{ll}$) and rapidity ($Y_{ll}$)
distributions for the final lepton pair with and without ISR corrections added using {\KK}MC-hh
(red) or switching to the LuxQED version of NNPDF3.1 (blue).}
\end{figure}

Angular distributions are of particular interest in the context of the measurement of the 
electroweak mixing angle at the LHC, in which {\KK}MC-hh is participating together with
other programs combining hadronic and electroweak effects, including POWHEG-EW\cite{PowhegEW}, 
MC-SANC\cite{MCsanc}, ZGRAD2\cite{ZGRAD2}, and HORACE\cite{HORACE}.
The Collins-Soper angle\cite{CollinsSoper} is the scattering angle in the CM frame of the final
lepton pair, given by
\begin{equation}
\cos(\theta_{\rm CS}) = {\rm sgn}(P_{ll}^z)
\frac{p_l^+ p_{\overline l}^- - p_l^- p_{\overline l}^+}{
\sqrt{P_{ll}^2 P_{ll}^+ P_{ll}^-}}
\end{equation}
neglecting masses,
with $P_{ll} = p_l + p_{\overline l}$ and $p^{\pm} = p^0 \pm p^z$. The initial-final interference
contribution is of particular interest for the angular distribution, since it is strongly 
dependent on the scattering angle. 

Fig. 2 shows the CS angle distribution generated in a 
{\KK}MC-hh run with $9\times 10^9$ muon events at an 8 TeV CM energy, using NNPDF3.1 
PDFs\cite{nnpdf3.1}, and without additional fermion cuts. 
Similar results are described in detail in Ref.\cite{KKMChhAFB}.  
The graph on the right of shows the $\cos(\theta_{\rm CS})$ full {\KK}MC-hh distribution (green) 
together with a version with IFI off (red), a version with both ISR and IFI off (black), and a 
version with IFI off, but ISR effects included by using a LuxQED\cite{LUXqed} 
version of NNPDF3.1 instead, NNPDF3.1-LuxQED\cite{nnpdfluxqed}
(blue). The shift due to ISR is $-0.4\%$ for both ways of accounting for ISR, to within 
$\pm 0.1\%$. The IFI correction is less than $10^{-4}$, but this effect 
increases for less inclusive cuts, as in the binned results for the forward-backward asymmetry
shown below.

\begin{figure}[ht]
\centering
\includegraphics[width=0.49\textwidth]{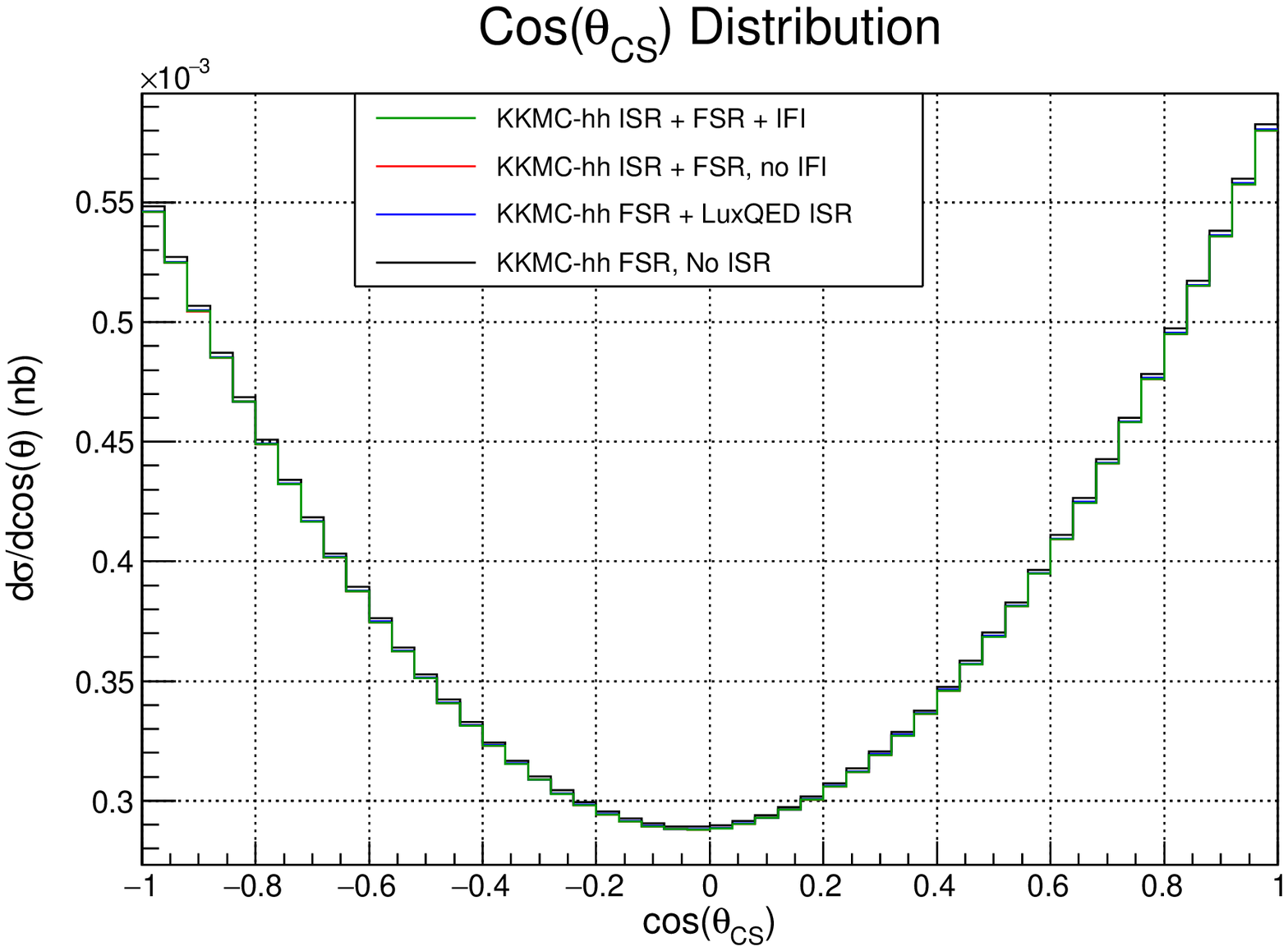}
\includegraphics[width=0.49\textwidth]{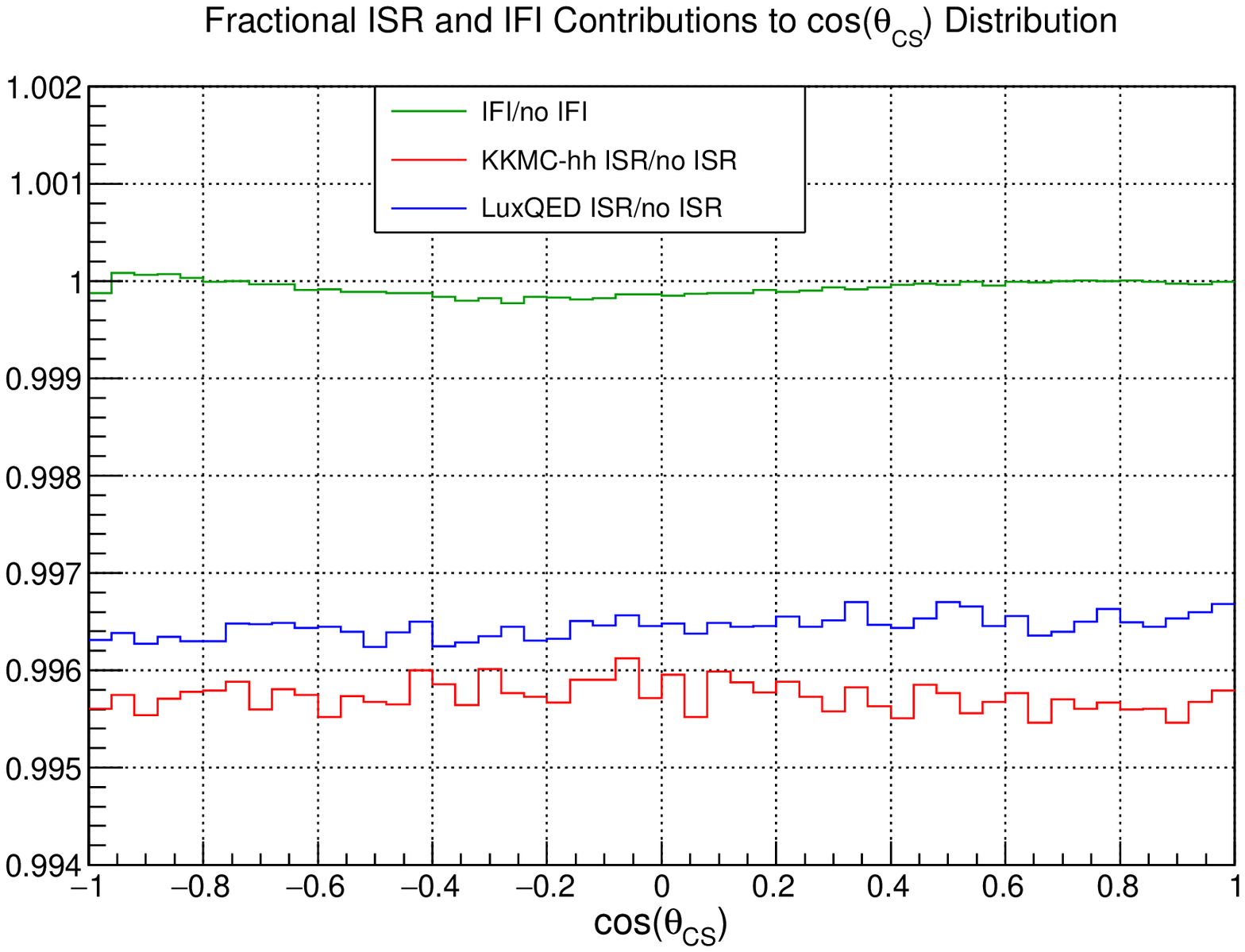}
\caption{The Collins-Soper angle distribution with full {\KK}MC-hh (green) compared to versions 
without IFI (red), without IFI and ISR (black), and without IFI but including ISR effects via a 
LuxQED version of the PDFs (blue). The graph on the right shows ratios with respect to a 
baseline result including only FSR corrections.
}
\end{figure}

The effect of the photonic corrections on the forward-backward asymmetry $A_{\rm FB}$ is 
shown in Fig. 3 as a function of $M_{ll}$ and Fig. 4 as a function of $Y_{ll}$. The IFI 
effect becomes much more pronounced in these binned results, while ISR is less 
significant, at least for cuts close to $M_{ll} = M_Z$. The right-hand graph of Fig. 3 shows that 
IFI has an effect (green) on the order of $0.1\%$ with a strong dependence on $M_{ll}$ in the 
vicinity of $M_{ll} = M_Z$.  The {\KK}MC-hh ISR effect is near zero and flat in the vicinity of 
$M_Z$, but with large statistical errors for larger $M_{ll}$. The ISR effect obtained by switching
from NNPDF3.1 to NNPDF3.1-LuxQED, however, shows a contribution small at $M_Z$ but larger
 elsewhere, and with a pronounced slope. ISR has negligible effect in Fig. 4, 
but IFI is strongly enhanced at high rapidities. 

\begin{figure}[ht]
\centering
\includegraphics[width=0.49\textwidth]{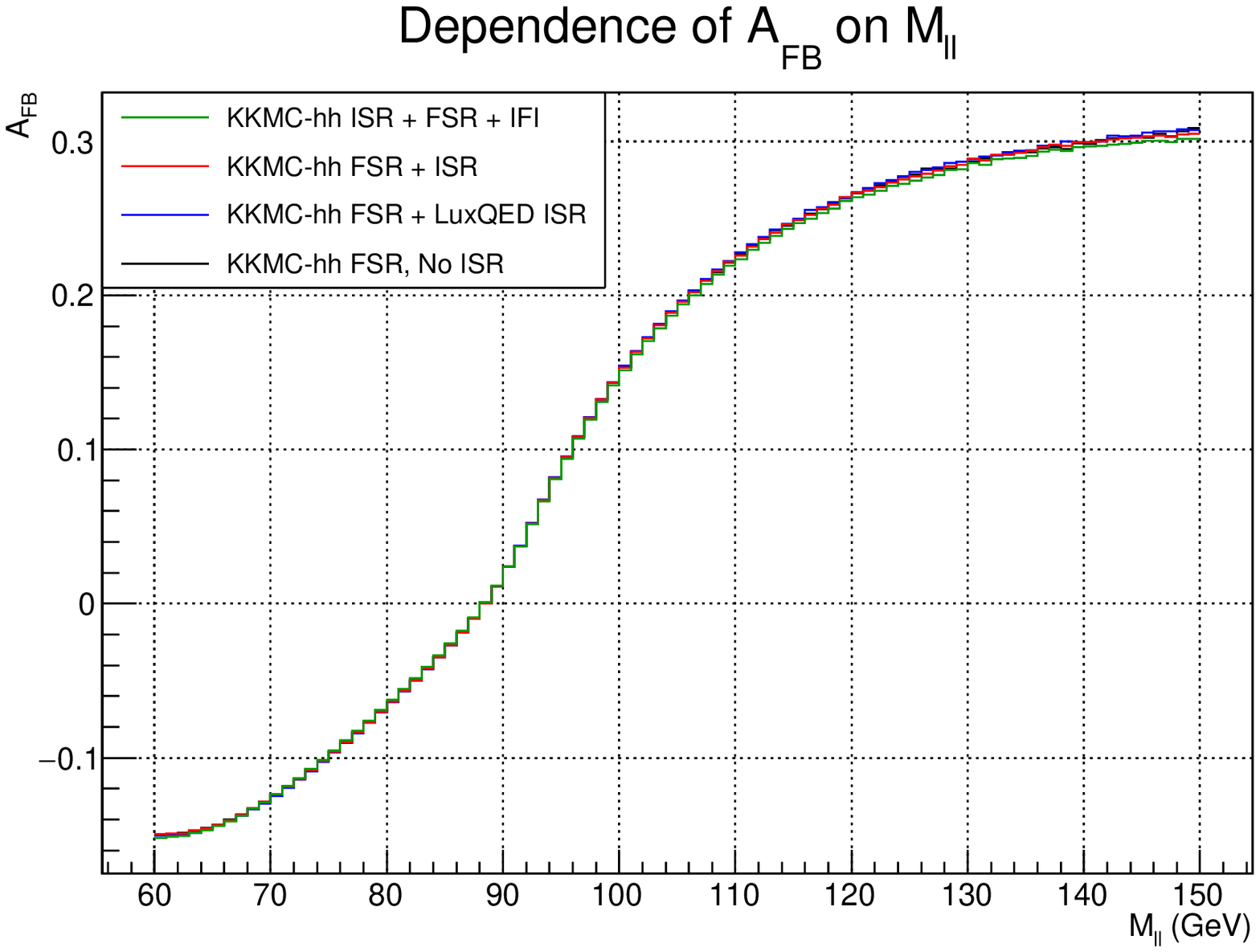}
\includegraphics[width=0.49\textwidth]{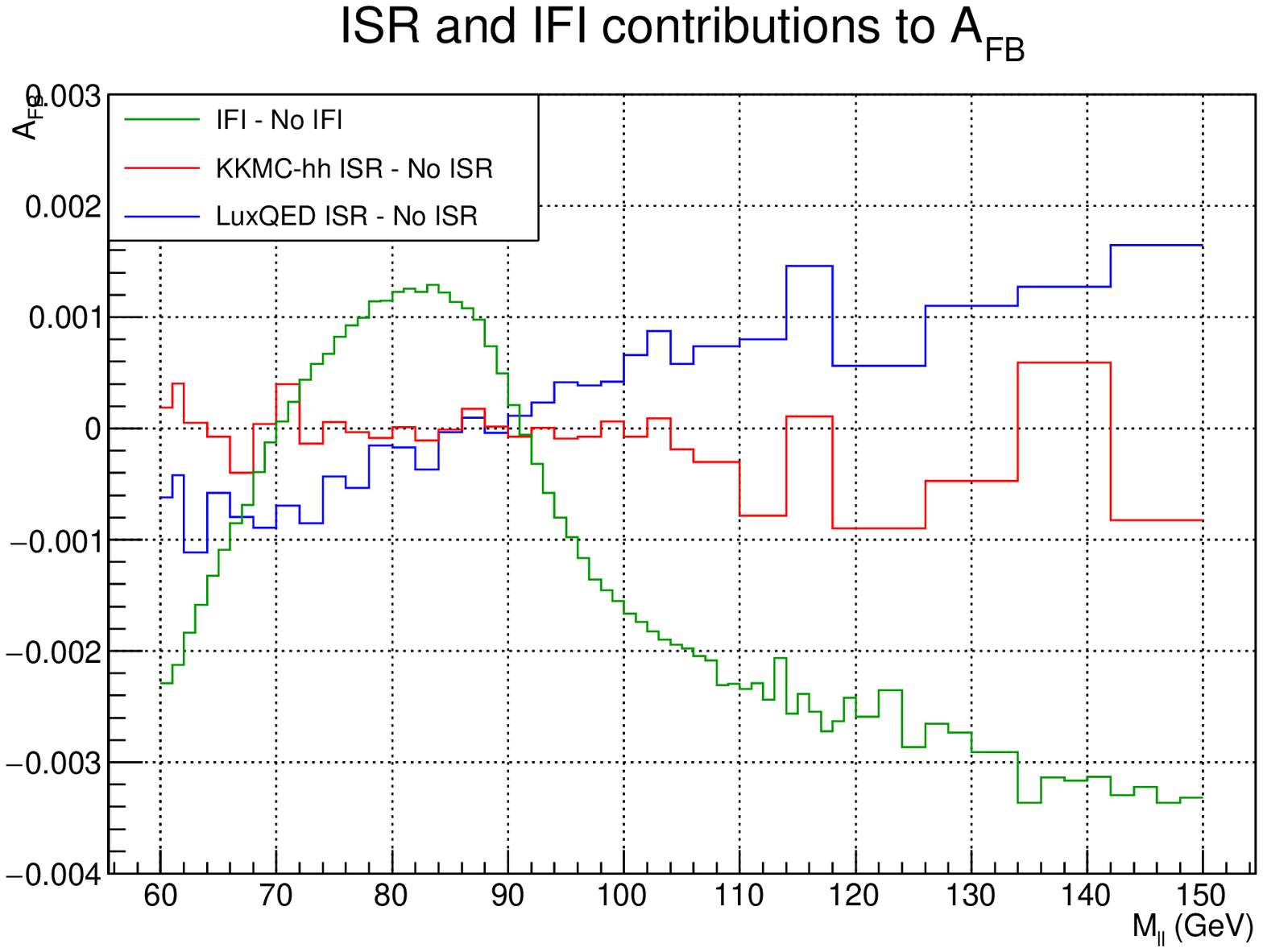}
\caption{The forward-backward asymmetry $A_{\rm FB}$ for full {\KK}MC-hh (green), as a function
of dilepton invariant mass $M_{ll}$,  compared to 
versions without IFI (red), without IFI and ISR (black), and without IFI but including ISR 
effects via a LuxQED version of the PDFs (blue). The graph on the right shows differences 
with respect to a baseline calculation with only FSR photonic corrections. 
}
\end{figure}

\begin{figure}[ht]
\centering
\includegraphics[width=0.49\textwidth]{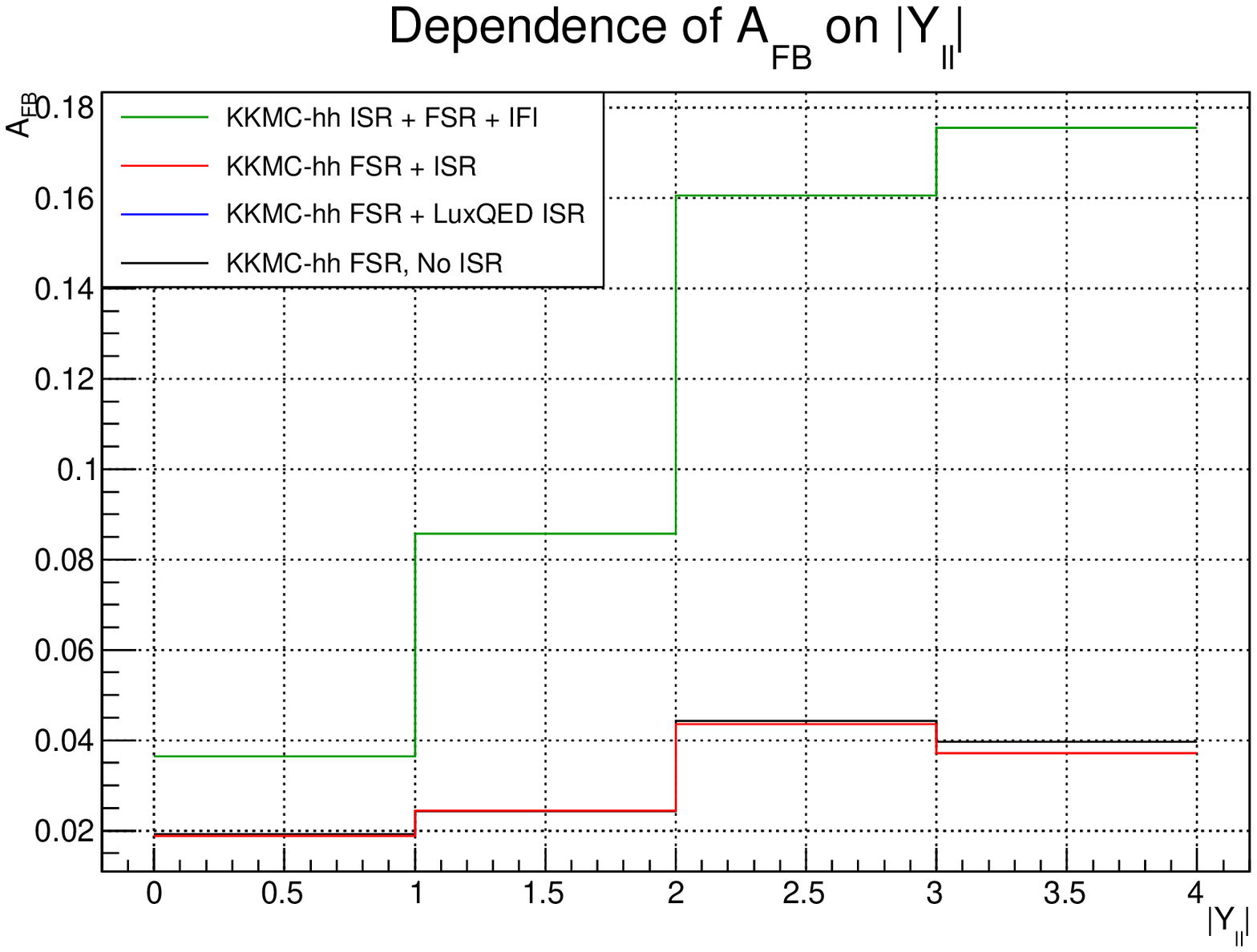}
\includegraphics[width=0.49\textwidth]{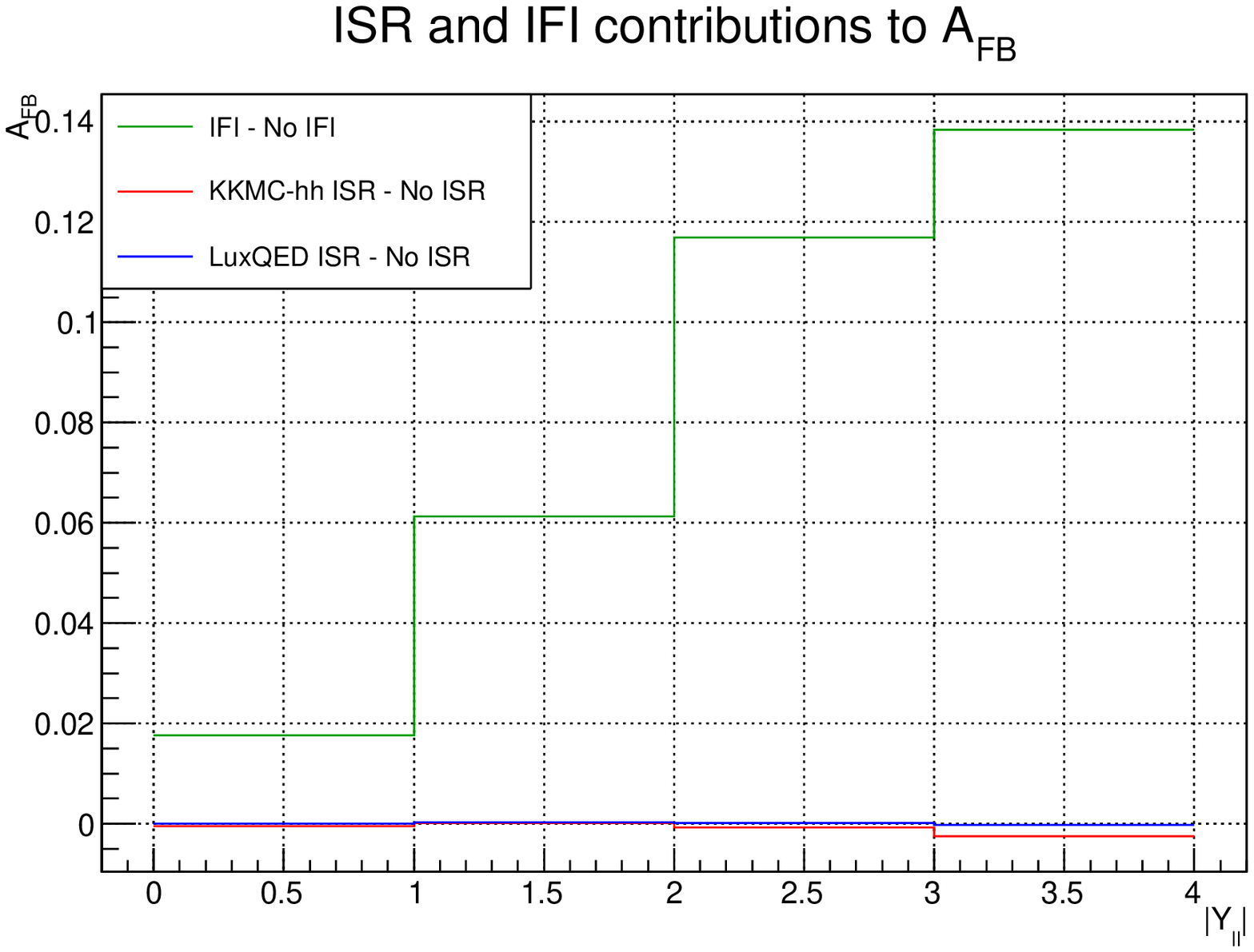}
\caption{The forward-backward asymmetry $A_{\rm FB}$ for full {\KK}MC-hh (green) as a function
of dilepton rapidity $Y_{ll}$, compared to 
versions without IFI (red), without IFI and ISR (black), and without IFI but including ISR 
effects via a LuxQED version of the PDFs (blue). The graph on the right shows differences 
with respect to a baseline calculation with only FSR photonic corrections.
}
\end{figure}

\section{Summary and Outlook} 
{\KK}MC-hh provides a precise tool for calculating exponentiated photonic
corrections to hadron scattering.  In particular, it can calculate 
contributions of ISR and IFI to the forward-backward asymmetry, 
which will be useful for determining the electroweak mixing angle from LHC data.
{\KK}MC-hh is particularly well suited to evaluating IFI due to its CEEX
exponentiation. 

While showered results were not presented in this note, they are possible both with an internal 
HERWIG6.5 shower, and by exporting events and applying an external shower.  This will allow 
addressing NLO QCD effects as well.  There has also been progress on a version of {\KK}MC-hh
which can be run to add electroweak corrections to previously-generated hadronic events.
{\KK}MC-hh is presently being transcoded entirely to C++.
This will facilitate compiling it with a current hadronic generator such as 
Herwig 7\cite{Herwig7} or KrkNLO\cite{krknlo}.

\section{Acknowledgments} 

S. Yost acknowledges support of The Citadel Foundation and the Institute of Nuclear Physics, 
IFJ-PAN, Krak{\'o}w, Poland, which provided computing resources for this project. S. Jadach
acknowledges support of National Science Centre, Poland Grant No. 2019/34/E/ST2/00457.
Z. W{\c a}s was supported in part by of Polish National Science Centre under 
decisions DEC-2017/27/B/ST2/01391.

\end{document}